\documentstyle[preprint]{aastex}
\def \CR		{{(42355)~2002~CR$_{46}$}}
\def \cR		{{CR$_{46}$}}

\def\TD		{{(29981)~1999~TD$_{10}$}} 	
\def\BU 		{{(33128)~1998~BU$_{48}$}}
\def\PJ 		{{(54520)~2000~PJ$_{30}$}}
\def\EE		{{(60608)~2000~EE$_{173}$}}
\def\OO		{{(87269)~2000~OO$_{67}$}}

\def \QT		{{(88611)~2001~QT$_{297}$}} 
\def \RZ  		{{(66652)~1999~RZ$_{253}$}}

\def \ref            {\noindent\hangindent0.5in\hangafter=1}

\begin{document}

\title  {Discovery of a Binary Centaur}

\author{Keith S.~Noll} 
\affil{Space Telescope Science Institute, 3700 San Martin Dr., Baltimore, MD 21218}
\email{noll@stsci.edu}

\author{Harold F.~Levison}
\affil{Dept.~of Space Studies, Southwest Research Institute, Ste.~400, 1050 Walnut St., Boulder, CO 80302}
\email{hal@boulder.swri.edu}

\author{ Will M.~Grundy}
\affil{Lowell Observatory, 1400 W.~Mars Hill Rd., Flagstaff, AZ 86001}
\email{grundy@lowell.edu}

\author{Denise C.~Stephens}
\affil{Johns Hopkins University, Dept.~Physics and Astronomy, Baltimore, MD 21218}
\email{stephens@pha.jhu.edu }

\newpage

\begin{abstract}  

We have identified a binary companion to \CR\ in our ongoing deep survey using the Hubble Space Telescope's High Resolution Camera.  It is the first companion to be found around an object in a non-resonant orbit that crosses the orbits of giant planets.  Objects in orbits of this kind, the Centaurs, have experienced repeated strong scattering with one or more giant planets and therefore the survival of binaries in this transient population has been in question.  Monte Carlo simulations suggest, however, that binaries in \CR -like heliocentric orbits have a high probability of survival for reasonable estimates of the binary's still-unknown system mass and separation.   Because Centaurs are thought to be precursors to short period comets, the question of the existence of binary comets naturally arises; none has yet been definitively identified.  The discovery of one binary in a sample of eight observed by HST suggests that binaries in this population may not be uncommon.

\end{abstract}


\newpage
\section{Introduction}

In 1977 a slow-moving, apparently asteroidal object, 1977 UB, was discovered in the course of a systematic survey of the ecliptic plane (Kowal~1977).  With the identification of several prediscovery images going back as far as 1895 (Liller et al.~1979a,b; Kowal et al.~1979) it quickly became apparent that this unusual object inhabited a chaotic orbit with close approaches to both Saturn and Uranus (Oikawa and Everhart 1979, Scholl 1979).  This object eventually received the minor planet designation 2060 and was named Chiron at the suggestion of Kowal.  He further suggested that similar objects also be named for mythological Centaurs.  The discovery of the Kuiper Belt, starting with 1992 QB$_1$ (Jewitt and Luu 1993), has led to the realization that objects like Chiron, in short-lived chaotic orbits, must have been recently perturbed from the more populous and more stable transneptunian population.  In addition, as many more objects have been found, the distinction between the Centaurs and transneptunian objects (TNOs), particularly the Scattered Disk objects (SDOs), has become blurred.  

A functional definition for Centaurs is that they are objects in the outer solar system on non-resonant, giant-planet-crossing orbits.  As a result of frequent encounters with the planets whose orbits they cross, they have dynamical lifetimes that are short, on the order of tens of millions of years.  Computation-intensive calculations are required to determine the orbital stability of any particular object (Horner et al.~2004a,b).  However, it is possible to identify classes of objects that are likely have unstable orbits.  The original definition of Centaurs included only those objects with semimajor axes, $a$, lying between the orbits of Jupiter and Neptune.  Holman (1997) has shown that, except for a small region of stability for low eccentricity orbits between 24 and 27 AU and for objects trapped in 1:1 resonance near the L$_4$ or L$_5$ Lagrange points of Jupiter and Neptune, objects in this region have dynamical lifetimes of 10$^8$ years or less with the current giant planet configuration.  Holman also notes that planetary migration would be likely to remove even the one small region of marginal orbit stability.  Thus, there is a very high likelihood that any object with a semimajor axis between Jupiter and Neptune will have an unstable orbit.  Most of the known non-Trojan objects with heliocentric semimajor axes between Jupiter's perihelion and Neptune's aphelion, $q_{Jupiter} < a < Q_{Neptune}$, are in non-resonant, giant-planet-crossing orbits that are unstable on timescales of order 10$^7$ years.  We note that there are a few objects that are not currently in planet-crossing orbits, such as 2001 XZ$_{255}$ which orbits between $q=15.43$ and $Q = 16.48$ AU.  Duncan and Levison (1997) found that such orbits are a natural feature in the chaotic evolution of objects undergoing interactions with the giant planets; the existence of a few such objects in the current inventory is not surprising and does not invalidate the assumption that all the known objects currently between Jupiter and Neptune have short dynamical lifetimes.  

Numerical simulations also show that non-resonant objects with $ q < Q_{Neptune}$ but $a > Q_{Neptune}$ are closely coupled to  cisneptunian Centaurs both in terms of their dynamical instability and by the fact that the semimajor axis of an individual object can move from inside Neptune to outside and vice versa through chaotic orbit evolution (Tiscareno and Malhotra 2003).  The Deep Ecliptic Survey (DES; Millis et al.~2002) adopts a broader definition of Centaurs that includes these objects as well: ``nonresonant objects whose osculating perihelia are less than the osculating semimajor axis of Neptune at any time during the integration'' (Elliot et al.~2005).  We note that a more consistent definition would also include objects with perihelia that come within Neptune's Hill radius of its osculating aphelion, although this is a minor quibble.  Other taxonomical schemes are possible (e.g.~Levison 1996), however, in this paper we adopt the DES definition of Centaurs.  In the DES scheme, the subject of this paper, \CR , is a Centaur on an unstable orbit that crosses the orbits of both Neptune and Uranus with $a = 38.1$ AU and $q = 17.5$ AU.

The first transneptunian binary (TNB) was identified in 1978 with the discovery of Charon (Christy and Harrington 1978), coincidentally contemporaneous with the discovery of the near-homonymously-named Chiron.  The subsequent discovery of a large population of TNOs and the more recent discovery of numerous bound systems profoundly changes the context in which we view binaries (see review by Noll (2006) for a recent summary).  Binaries and more complex bound systems can no longer be regarded as low probability outcomes of statistically unlikely events, but must instead be explained as a natural consequence of small body evolution in the outer solar system.  Sufficient data on binary frequency have now accumulated that differences in the binary frequency between different dynamical populations are beginning to be identifiable.  Stephens and Noll (2006) have identified an enhanced occurrence of binaries in the cold classical population compared to other dynamical groups of TNOs.  This difference has helped reinforce speculation as to the instability of binaries in populations that have had strong interactions with giant planets.  Centaurs have experienced repeated strong scattering events with one or more giant planets.  It is an open question whether typical scattering events are capable of disrupting a significant fraction of weakly bound systems, a question perhaps best addressed by both observation and theory.

In order to obtain the least biased statistics on the occurrence of binaries in different subpopulations of small bodies in the outer solar system we have undertaken a survey using the High Resolution Camera (HRC), a component of the Advanced Camera for Surveys (ACS) on the Hubble Space Telescope (HST).  Our program will observe a subset of 250 possible targets distributed over all known dynamical classes in a snapshot program that is active from July 2005 through June 2007.  Observations with the HRC are much more sensitive to binaries than previous searches both in terms of the limiting magnitude and in the angular resolution of the instrument.  The program includes an emphasis on currently undersampled populations including the Centaurs.  In this paper we describe the discovery of a binary companion to \CR , the first known binary Centaur.

\section{Observation and Analysis}

Observations of \CR\ were made with ACS/HRC on 20 January 2006, 09:50 - 10:14 UT.  The HRC was used in a non-standard configuration with the $Clear$ filters in both filter wheels rotated into the optical path.  The use of two $Clear$ filters results in a factor of two increase in throughput compared to the widest available standard filter at the cost of a slight degradation in image quality (Gilliland and Hartig 2003).  The bandpass is determined by the detector response and is equivalent to a filter centered at $610\,$nm with a FWHM of $520\,$nm.  Four $300\,$sec exposures of \CR\ were obtained.  The telescope was dithered between each exposure using the standard ACS/HRC box dither pattern.  The use of dithers ensures that fixed-pattern defects such as hot pixels can be removed when the exposures are coadded.  By obtaining four exposures we are able to eliminate essentially all cosmic rays when the individual exposures are combined.  

The telescope tracked the motion of \CR\ during the exposures and corrected for parallax introduced by HST's orbital motion.  Observations for all objects in our program have been scheduled with a restriction that the target and background sources must have relative motions at a minimum rate of 5$\times 10^{-5}$ arcsec/sec to ensure that we can discriminate between the target and background sources.  All of the targets observed to date have been successfully acquired in the 29$\times$26 arcsec field of view of the HRC's 1024$\times$1024 CCD array.

Data were reduced using the standard CAL ACS pipeline calibration.  For moving targets, we start our analysis with the dark-subtracted, flat-field corrected $*flt.fits$ image files; the $*drz.fits$ combined image file from the standard pipeline is not useable for moving targets.  The binary companion to \CR\ is apparent in each of the individual flat-fielded images without requiring any further processing.  However, in order to produce an image that is free of cosmic rays, hot pixels, and that has the best signal-to-noise ratio ($S/N$), we remove geometric distortion and combine the images using the $multidrizzle$ task (version 2.6.7) in the $STSDAS$ package running in $PyRAF$. 

To combine moving target observations we must add a nonstandard step to the usual $multidrizzle$ work flow.  Because the target is tracked, the coordinate keywords in each exposure are different, reflecting the different starting position of each integration.  In order to remove the target-tracking motion of the telescope that would be misinterpreted by $multidrizzle$, we force the world coordinate header keywords $CRVAL1$ and $CRVAL2$ for each of the four exposures to be equal to the value at the start of the first exposure.  Each image is centroided and shifts are computed using the $tweakshifts$ command.  The computed shifts are used make a final combination of the four exposures.  The result of this process is shown in Figure~1.  The signal in the peak pixel has $S/N=2700$ and the standard deviation in the background is 0.0078 counts/sec.  The 3 sigma, peak-pixel detection limit corresponds to a source of $V=27.7$ mag.  Note that peak-pixel $S/N$ can be improved by a factor of $\sim$3 when the data are fit to a point-spread-function (PSF).  The level of performance observed is in excellent agreement with predictions based on the ACS exposure time calculators and confirms the success of data processing.

We measured the separation by iteratively fitting a PSF to each of the two components of \CR\ in each of the four exposures and in the combined exposure.  The iterative PSF fitting is described in detail in Stephens and Noll (2006).  We found a separation of 4.36$\pm$0.08 pixels, equal to an angular separation of 0.109$\pm$0.002 arcsec.  The secondary was oriented at an angle of 226.8$\pm$0.8 degrees East of North.  PSF fitting also yielded scaling factors that we used to derive relative photometry of the two components.  We found the secondary to be fainter than the primary by 1.47$\pm$0.04 magnitudes.  The measured magnitude difference corresponds to a diameter ratio of $d_2/d_1 \approx 0.5$, assuming both components have the same albedo.  Computing standard magnitudes is more difficult because of the extremely wide effective bandpass of the $Clear$ filter and the uncertain colors of the objects; we do not attempt it here.

\section{Discussion}

\CR\ was discovered by the NEAT team at Palomar and the heliocentric orbit was refined soon thereafter (McNaught et al.~2002; Stoss and Marsden 2002).  At the time of our HST observation the distance from Earth to \CR\ was $16.675\,$AU; a plan view of the orbit is shown in Figure~2.  The projected separation of the two components on the plane of the sky was 1330$\pm130\,$km.  The absolute magnitude of \CR\ is $H_V$ = 7.65$\pm$0.01 (Tegler et al.~2003).  Optical colors measured for this object are $B-V = 0.74\pm0.02$ and $V-R = 0.52\pm 0.01$ (Tegler et al.~2003); colors that are typical of the relatively ``neutral'' group of Centaurs and TNOs.  Colors measured by Peixinho et al.~(2004) differ slightly, but are generally within the quoted uncertainties.  Ortiz et al.~(2003) placed an upper limit of 0.15 mags on any possible lightcurve from the unresolved binary.  The albedo of the unresolved binary was measured by Stansberry et al.~(2005) using the Spitzer Space Telescope; they reported $p=0.08-0.12$.  For assumed albedos of $p=0.1$ for each component and $H_V$ = 7.65 for the pair, we calculate a total equivalent diameter of 125 km.  For a 2/1 diameter ratio we calculate component diameters of 112\,km and 56\,km for the primary and secondary respectively.  

An interesting question to ask for any binary is how tightly bound the system is.  There are two ways to answer this question, in terms of the total binding energy of the system, and in terms of the stability of the system relative to perturbations from other bodies.  The binding energy depends on the separation of the secondary from the primary, $a_B$, and on the system mass.  By assuming that densities of TNOs are similar (an admittedly shaky assumption), the quantity $r_p^3/a_B$, where $r_p$ is the radius of the primary, can be used to infer the relative binding energies of bound systems.  More frequently, however, the related quantity $a_B/r_p$ is tabulated, a convention we follow here.  Based on a single observation we cannot constrain the semimajor axis of the binary, $a_B$, beyond the statement $a_B > (1330/2)\,$km which would be true for the limiting case of an orbit with eccentricity $e$ = 1 seen face on.  A better guess for the semimajor axis is twice the observed separation, $2\times 1330 \approx 2700\,$km.   Using this, we estimate $a_B/r_p = 48$.  This is comparable to the value of $a_B/r_p = 56$ found for \RZ\ and significantly less than for any other TNB with a measured orbit except for the Pluto/Charon system (Noll 2006) indicating that the \cR\ system is more tightly bound than most other known binaries.    To estimate the system stability relative to third body perturbations we assume a density near $1\,$g~cm$^{-3}$ giving a Hill radius, $r_H$, at the current heliocentric semimajor axis of $38.1\,$AU on the order of $300,000\,$km (note that the instantaneous Hill radius scales with distance from the Sun).  Again using twice the currently observed projected separation, $s$, in place of the still-to-be-determined semimajor axis, $a_B$, we find $2s/r_H \approx$ 0.01.  Separations on the order of few percent of the Hill radius are fairly typical in binaries in the transneptunian, Main Belt, and Near Earth populations (Noll 2006) and, in this sense, \CR\ does not appear to be unusual.

\subsection{Frequency of Binary Centaurs}

An important question in the study of binary systems in the transneptunian and related populations is whether and how the fraction of binaries in different dynamical populations vary.  Existing data show an increased fraction of binaries in the cold classical disk compared to an aggregate of the remaining populations (Stephens and Noll 2006).  A plausible, but unverified postulate is that populations that have experienced disruptive scattering events will have a smaller fraction of binaries than populations that have had quiescent dynamical histories.  Centaurs are very short-lived compared to the age of the solar system and during their lives undergo many encounters with giant planets.  The Centaurs, following this line of reasoning, might be expected to have the lowest fraction of surviving binaries as a result of repeated scattering encounters with giant planets.  However, while there has been informal speculation on this topic, the observations needed to address it quantitatively and/or the numerical modeling required for a sound theoretical prediction have not been completed.   

The DES currently lists 65 objects as Centaurs (Elliot et al.~2005) using their broad definition.  None of these has been found to be binary from existing groundbased observations.  To gauge whether this lack of binary discovery is surprising, it is most useful to make a comparison with binaries found in the general TNO population.  Of the approximately 1000 TNOs discovered from ground-based observations, 7 have been identified as binaries from non-AO ground-based images.  If the same fraction of ground-detectable binaries existed in the Centaur population, we would expect 1 binary, at most, to have been detected in that sample.  We note that the groundbased survey of Schaller and Brown (2003) failed to identify any binaries in a sample of 150 TNOs, again consistent with the overall infrequency of wide binaries.  The smaller heliocentric distance of Centaurs favors the detection of similarly wide binaries compared to TNBs by a factor of order two; but this is insufficient to change our conclusion that the current lack of ground-detected Centaur binaries does not constrain Centaur binaries to be less frequent than TNBs.

High resolution observations with HST and with adaptive optics systems from the ground have found the majority of known TNBs (Noll 2006), largely due to the fact that component separations are typically much less than 1 arcsec.  Eight Centaurs have been observed by HST using four different imaging instruments, the Wide Field Planetary Camera 2 (WFPC2), the Space Telescope Imaging Spectrograph (STIS), the Near Infrared Camera and Multi-Object Spectrometer (NICMOS), and the ACS/HRC (Table 1).  While this is far from being a homogeneous sample in terms of resolution or depth, it is nevertheless far less heterogeneous than ground-based data and is a reasonable starting point for estimating the binary frequency in Centaurs.  

In order to perform the most sensitive possible search for faint companions, we combined all available data for each object with $multidrizzle$ using the same steps as described for \CR .  This analysis step necessarily involved some compromises.  The data were obtained in multiple filters, the F555W, F675W and F814W for WFPC2 and the F110W and F160W for NICMOS.  Because we are combining data with different intrinsic point-spread-functions (PSF) we lose the ability to extract unresolved binaries from PSF-fitting (e.g. Stephens and Noll 2006) but increase our ability to detect faint resolved companions.  The angular detection limit for data processed in this way is then given by the standard Nyquist criterion.  For WFPC2 the limit is 0.2 arcsec, for NICMOS it is 0.15 arcsec.  We note, as a caveat, that the resolution of the combined image also depends on the accuracy of the centroiding of the individual frames and the calculated shifts used for alignment.  For low $S/N$ data (\PJ ) or for undersampled data (\EE\ and \OO ) additional systematic errors may accrue.

STIS data were taken in fixed target mode; the Centaurs move by several pixels during each of the integrations.  This motion is due mostly to HST parallax and is therefore nonlinear both in direction and magnitude in the course of an orbit.  We combined a subset of eight frames for each of the objects, choosing the integrations with the smallest degree of blurring.  The angular resolution of the combined image is complex and is a function of position angle.  However, we estimate that a companion at a distance of 0.2 arcsec or greater would be resolved.

For all of the objects we have estimated how much fainter an object could have been detected in the combined image.  In all cases but one, the combined image has a smooth background with few artifacts except for residuals from extended sources that are incompletely removed.  The apparent motion of \OO\ relative to stars in the field was small enough that the stars were not removed by the $multidrizzle$ cosmic ray step (essentially a median filter).  However, the background outside these sources is flat and clean.  The standard deviation of the background was determined by using $imstat$ on clean subsections of the images.  We estimated the detectability limit for a faint source to be three times this standard deviation in the peak pixel.  The ratio of the peak pixel in the observed Centaur to this detectability limit, expressed in astronomical magnitudes, is listed in Table~1.  This depends on the brightness of the Centaur as well as the depth of the integration and varies widely.  It is worth noting that with the exception of the two newly-detected companions observed around Pluto (Weaver et al.~2006),  all other known binaries have magnitude differences of $\sim 4.5$ magnitudes or less (Noll 2006, Brown et al.~2006); all but two of our data sets probe to 4.5 magnitudes or deeper.  Most TNBs, 17/25, have $r_s/r_p > 0.5$ and would have been detectable in all of our data sets.

The detection of one binary in the HST sample of eight Centaurs corresponds to a binary frequency of 13$\pm {19 \atop 5}$\%  at separations of greater than 0.1-0.2 arcsec for companions within a factor of 3 of the brightness of the primary.  Upper limits for significantly fainter companions at these separations can also be derived from this sample based on the data in Table~1.  We can compare to other dynamical classes measured from NICMOS data; Stephens and Noll (2006) found binary rates of 22$\pm {10 \atop 5}$\% for objects in the cold classical belt and 11.5$\pm {9 \atop 4}$\% for SDOs.  Within the large uncertainties, the fraction of binary Centaurs cannot be distinguished from either the SDOs or the Cold Classicals.  It appears unlikely, however, that Centaur binaries are significantly less frequent than binaries in the Scattered Disk.

\subsection{Survival of the \CR\ Binary}

It is possible to estimate whether a binary system like \CR\ could survive as it evolved from an orbit originally in the Scattered Disk into the orbit in which it is currently found.  Such an estimate requires two distinct calculations.  First, we must determine, at least statistically, the trajectory that the object followed after it left the Scattered Disk, paying particular attention to close encounters with the planets.  Second, we must determine what kind of planetary encounters will lead to the tidal stripping of the binary.

Fortunately, we can accomplish the first task by employing a set of numerical orbit integrations presented in Levison et al.~(2006a, hereafter L06).  L06 followed the dynamical evolution of 2200 objects initially in the Scattered Disk under the gravitational effects of the Sun and all the planets (except Mercury, Pluto and other large TNOs).  Each trajectory was followed until the object was either ejected from the solar system, impacted the Sun or a planet, or reached a semimajor axis of $1000\,$AU, where it was assumed it would enter the Oort cloud.  During the simulation, L06 kept track of the close encounters with the planets.

For our purposes, we will consider any object on an orbit with semimajor axis $37<a<39\,$AU, perihelion distance $16<q<18.5\,$AU, and inclination $1<i<3^\circ$ to be on an orbit similar to that of \CR .  Using this definition, 126 of L06's 2200 objects evolved into \cR -like orbits sometime after they left the Scattered Disk, roughly 6\%.  We refer to these 126 objects as our {\it fictitious \cR 's}.  The paths they took varied greatly from object to object because their orbits are chaotic.  For example, they took between $2\,$Myr and $1.4\,$Gyr between the first close encounter with Neptune and arrival into a \cR-like orbit for the first time.  The median time for this transition was $48\,$Myr.

There is a large variation in the pathways that the fictitious \cR 's took, as mentioned above, and thus it is not possible to define a `typical trajectory'.  Nevertheless, it is still educational to examine an example.  One such example is presented in Figure~3.  The solid curve in the figure shows the evolution of the semimajor axis of one of the fictitious \cR 's as a function of time.  The two dotted curves show the evolution of the object's perihelion and aphelion distances.  Initially, the object had a semimajor axis of $44.5\,$AU, but its eccentricity was large enough that it could suffer close encounters with Neptune.  Indeed, for the first $\sim\!6\,$Myr, the dynamical evolution of this object is driven by Neptune encounters.  Between $6.5$ and $10.6\,$Myr the object becomes temporarily trapped in Neptune's 4:5 mean motion resonance.  This resonance lifts the object's perihelion distance away from Neptune, thereby supplying a short-lived reprieve from encounters.  At $10.6\,$Myr encounters start again, and at $12\,$Myr the object is scattered by Neptune onto an orbit with $a=22.5\,$AU, $q=14.5\,$AU and $Q=30.5\,$AU (see Levison and Duncan~1997 for a description of this behavior).  After the scattering event at 12 Myr, encounters with Uranus drive the evolution of this object, and at $17\,$Myr it evolves onto a \cR -like orbit.

In their 2003 paper, Tiscareno and Malhotra (serendipitously) computed the orbit for a test object with initial conditions identical to \CR\ (Tiscareno and Malhotra 2003, Fig.~4).  Their test object orbit evolved with a roughly constant perihelion distance but increasing semimajor axis until, at 34 Myr, the object was lost to the inner solar system.  Because of the chaotic nature of Centaur orbits, this result is not unique, but is again instructive as an example of the kind of behavior that is possible.

The behaviors described above illustrate the importance that close encounters with the giant planets have in creating objects on \cR -like orbits. Here we define `close' as an encounter in which the objects passes within the Hill sphere of a planet.  The object shown in Figure~3 suffered 147 close encounters with Neptune and 41 close encounters with Uranus over the $18\,$Myr time-frame shown.  Not surprisingly, for our ensemble of fictitious \cR 's, close encounters with Neptune are the most prevalent.  These objects suffered between 39 and 1657 (median 288) such events on their way to a \cR -like orbit.  Encounters with Uranus numbered between 0 and 363, with a median of 57 (only 1 of our 126 fictitious \cR 's did not suffer an encounter with Uranus at all). Interestingly, 9\% of our objects suffered at least one encounter with Saturn and 3\% with Jupiter {\it before} evolving onto \cR -like orbits.  

Whether  \CR\ survives as a binary through this evolution depends, to a large extent, on the strength of the encounters, which, in turn, depends on the closest approach distances.  In Figure~4 we present the cumulative probability that one of our objects suffered an encounter within a given distance of one of the giant planets on its way to a \cR -like orbit.  Half of our fictitious \cR 's suffer an encounter no closer than $\sim\!0.1\,$AU and 90\% of them do not get closer than $0.03\,$AU.

The next step in our investigation is to determine whether these encounters will disrupt the binary.  The exact outcome of an encounter depends on many parameters including the close approach distance ($D$), the total mass of the binary pair ($m_{tot}$), the mass of the planet ($M_p$), and the details of the binary's mutual orbit.  However, Agnor and Hamilton~(2005) found that for nearly parabolic encounters disruption becomes common for 
\begin{equation} 
D<r_{td} \equiv a_B \left( \frac{3 M_p}{m_{tot}}\right)^{1/3}. \end{equation}

We do not know, yet, the binary's orbit and, therefore, must estimate $a_B$ and $m_{tot}$ if we are to to gauge the likelihood that \cR\ would have survived as a binary to its current heliocentric orbit.  It is unlikely (between 14\% and 19\% depending on the eccentricity) that the semimajor axis of the binary is larger than twice the $1330\,$km projected separation; we have adopted $a_B=2700\,$km as a conservative guess for this calculation.  We note that subsequent observations of \CR\ do not contradict this guess (Grundy et al., in preparation).   We estimated above that \CR\ has a combined equivalent diameter of $125\,$km, which, for an assumed density of $1\,$g$\,$cm$^{-3}$, implies a system mass of $10^{21}\,$g.  Under these assumptions, we find that $r_{td}$ ranges from $0.0067\,$AU (for Neptune) to $0.018\,$AU (for Jupiter).  The upward arrows in Figure~4 show the values of $r_{td}$ for the four giant planets.

Comparing the curves in Figure~4 to the values of $r_{td}$ for the various planets, we estimate that there is a 95\% chance that the \cR\ binary would have survived its transition from the Scattered Disk to it current orbit if $a_B\!\sim\!2700\,$km.  Half of the threat to the binary comes from Uranus and half from Jupiter.  We can turn the calculation around and determine what values of $a_B$ are vulnerable to disruption.  Using the data in Figure~4 and assuming $m_{tot} = 10^{21}\,$g, we find that an object with $a_B\!\sim\!23,000\,$km has only a 50\% chance of survival.  If we assume that the albedo of \CR\ is 1 rather than 0.1, i.e. the objects are the minimum possible diameters and therefore lower in mass, the 50\% survival semimajor axis falls to $7300\,$km.  

Finally, it is interesting to question whether \CR\ might be expected to have a small semimajor axis relative to the primary diameter because any binary that was less tightly bound would not have survived.  This query can be addressed by calculating $r_{td}$ for the known TNBs and comparing these values to the curves in Figure~4.  For the 8 TNBs, excluding Pluto, with measured orbits (as summarized in Table~6 of Noll~2006), we find that $r_{td}$ with respect to Uranus ranges from 0.007 to $0.09\,$AU.  Even in the most vulnerable case, \QT , there is a 68\% chance that the binary would have survived.  Thus, we can conclude that \CR 's semimajor axis is not strongly bounded by a dynamical selection effect.

\subsection{Where Are the Binary Comets?}

Jupiter family comets (JFCs) have been shown to derive from a source population on prograde, low inclination orbits at larger heliocentric distances and were, indeed, the first evidence pointing to the existence of a transneptunian population, particularly the SDOs (Duncan et al.~2004 and references therein).  Centaurs are a natural feature of this model as some will transition from large to small heliocentric orbits under the influence of the giant planets.  If binary Centaurs exist at any appreciable fraction, the next natural question is whether there are binary JFCs.  This same question could be asked for Halley-type comets since their source may be SDOs (Levison et al.~2006b) and Scattered Disk binaries have been identified (Noll 2006; Noll et al.~2006).  Indeed, in view of the fact that binaries seem to be a common phenomenon in various small body populations, the question could as well be applied to Oort-cloud comets. 

Excluding objects that have been observed to disintegrate into one of more pieces, there has been, to date, only one comet of any kind with even tentative claims for binarity.  Marchis et al.~(1999) and Sekanina (1999) both claimed evidence of a possible binary nucleus for the large Oort-cloud comet Hale-Bopp.  Sekanina identified a companion with a diameter 0.4 times that of the primary in HST WFPC2 images processed to remove light from the coma.  Marchis et al.~(1999) pointed out both the double peaked coma they obtain after image restoration and morphological features of the jets as possible evidence of a binary.  However, Weaver and Lamy's (1999) analysis of Hale-Bopp employing some of the same HST data used by Sekanina found a possibly smaller diameter for the nucleus (30-70 km vs.~70 km) and reported no evidence of a second component.  

In the absence of a direct detection of a binary nucleus, we can ask what other lines of indirect evidence might point to the existence of binary comets.  Comet fragmentation is a well-known phenomenon that has been observed in many comets (Boehnhardt 2004).  Historically, these events have been interpreted as physical breakup of a single nucleus.  We pose a speculative question: could some of the observed split comets be due to a preexisting binary?   Given the predominance of similar-mass binaries in the transneptunian population (Noll 2006), an expected outcome of chaos-assisted capture (Astakhov et al.~2005), the most likely candidates in this speculative scenario would be those comets that appear to split into nearly equal sized components without the obvious influence solar heating as a driver of the breakup.  An example of a split comet that might be amenable to interpretation as a preexisting binary is the comet pair C/2002 A1 and C/2002 A2 (LINEAR).  These two objects are on very similar Centaur-like orbits with semimajor axes of 18 AU, perihelia at 4.7 AU and aphelia at 31 AU and nearly identical eccentricities, inclinations and magnitudes (Spahr et al.~2002).    Sekanina et al.~(2003) used their comet fragmentation code and repeated astrometric observations to infer that the pair split non-tidally at a distance of $\sim$22 AU from the Sun in the recent past.  However, they did not consider the possibility of a preexisting binary in their model and it is unclear what observable differences, if any, one might expect between a single, internally fractured progenitor and a tight binary.    

An important ancillary issue must be mentioned.  The nuclei of JFCs are typically much smaller than known Centaurs or TNOs (Lamy et al.~2004).  The dependence of binary fraction on diameter is currently unknown and will require significant investments of observing time to answer for any of the small-body populations in question.  It is conceivable, however, that there could be a dependency that would weaken any comparisons between tens-of-km-scale Centaurs and km-scale JFCs.  Nonetheless, given the apparent existence of binary Centaurs, more observational and theoretical attention to the question of binary comets is certainly in order.

\section{Conclusions}

\CR\ is a binary system, the first binary Centaur to be identified.  \CR 's orbit crosses the orbits of both Uranus and Neptune and it is, therefore, in a dynamically unstable orbit.  Numerical simulations show that, on average, an object on its way to a \cR-like orbit will have had hundreds of close encounters with giant planets with a 50\% chance of coming as close as  D = 0.1 AU to one.  This system is, then, an excellent empirical test of the proposition that close planetary encounters can disrupt binaries.  The existence of an object like \CR\ rules out the extreme formulation of the proposition that all binaries on giant-planet-crossing orbits will be disrupted.  The separation of the components of the \CR\ system compared to the size of the primary is smaller than for most known TNBs; an indication that this system is tightly bound.  In order to test the weaker formulation of the binary-disruption postulate, i.e. that some fraction of planet-crossing binaries will be disrupted leaving only the most tightly bound systems, many more Centaurs will need to be observed at high angular resolution.

\section{Postscript}

Since this manuscript was prepared and reviewed, four additional Centaurs have been observed in our ongoing HST observing program.  Three are single objects ( (120061) 2003 CO$_1$, (55576) Amycus, and (83982) Crantor).  One more Centaur, (65489) 2003 FX$_{128}$, is binary.  The detection of another binary in an unstable giant-planet-crossing orbit strengthens the conclusions of this paper.  With the addition of these four Centaurs, the statistics for Centaur binaries stands at 2 out of 12.   The details of these recent observations will be covered in future publications.

\acknowledgements {This work is based on observations made with the NASA/ESA Hubble Space Telescope. These observations are associated with program \#~10514.  Support for program \#~10514 was provided by NASA through a grant from the Space Telescope Science Institute, which is operated by the Association of Universities for Research in Astronomy, Inc., under NASA contract NAS 5-26555.  HFL is grateful the PG\&G and Origins for continuing support.}

\begin{figure}
\includegraphics[totalheight=0.6\textheight,angle=0]{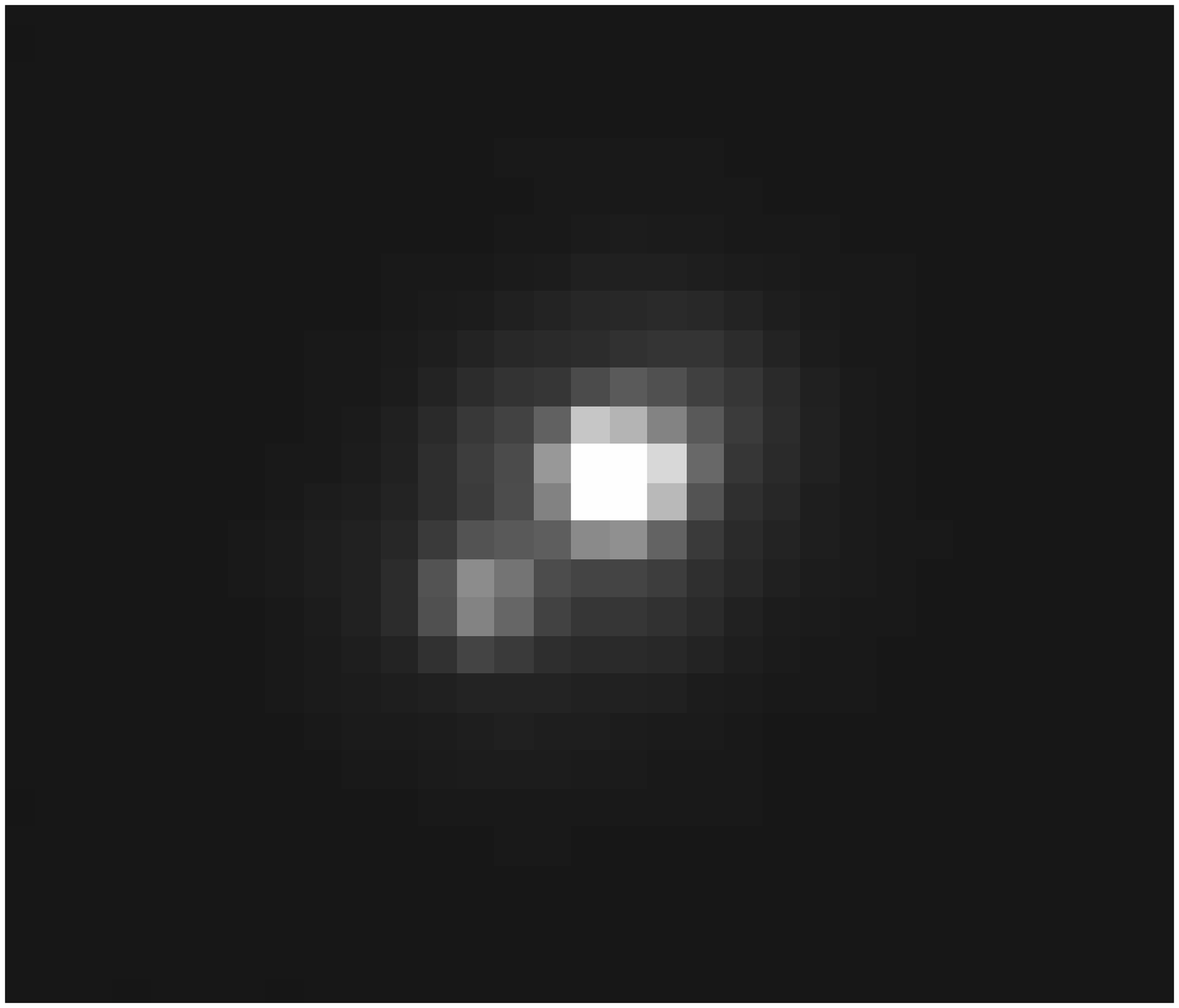}
\caption {Combined image of \CR\ obtained with the ACS/HRC is shown.  Four separate 300 sec integrations have been combined as described in the text.  The secondary is clearly resolved to the lower left of the primary.  The pixels in this image are 25 milliarcsec on a side; only a small portion of the full image is shown for detail. The secondary lies 0.109$\pm$0.002 arcsec from the primary at a position angle of 226.8$\pm$0.8 degrees East of North.   North, in this image, lies approximately to the right.}

\label{fig1}
\end{figure}

\begin{figure}
\includegraphics[totalheight=0.6\textheight,angle=0]{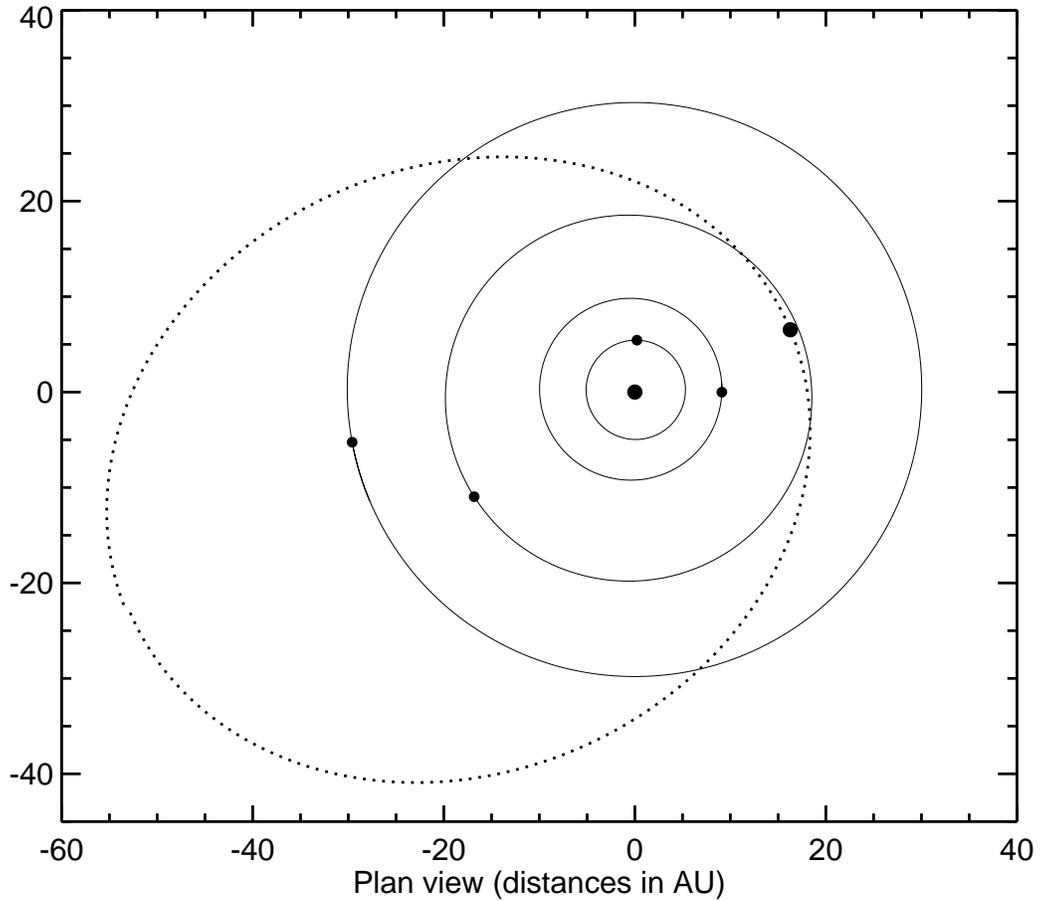}
\caption {A plan view of the orbit of \CR\ compared to the orbits of the gas-giant planets.  Dots show the location of each object on 1 March 2006.  The orbit of \CR\ crosses both the orbits of Neptune and Uranus.  It is not in an identified resonance with either planet, will have close encounters with both, and will eventually be scattered out of this region of the solar system.  }

\label{fig2}
\end{figure}

\begin{figure}
\epsscale{0.8}
\plotone{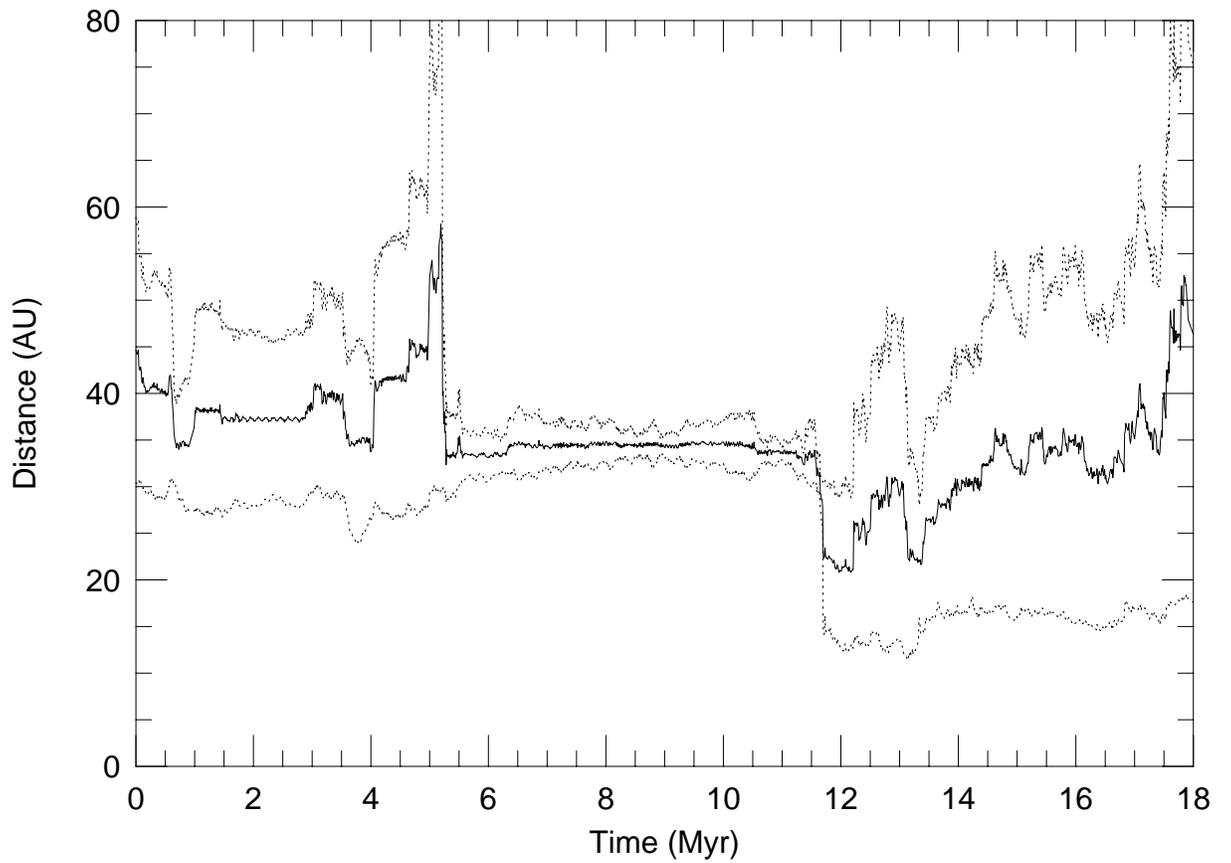}
\caption{The semimajor axis (solid), perihelion distance (lower dotted), and aphelion distance (upper dotted) of a \CR-like test object as it evolves over 18 Myr through repeated encounters with giant planets is shown.}
\label{fig_aqQ} 
\end{figure}

\begin{figure}
\epsscale{0.8}
\plotone{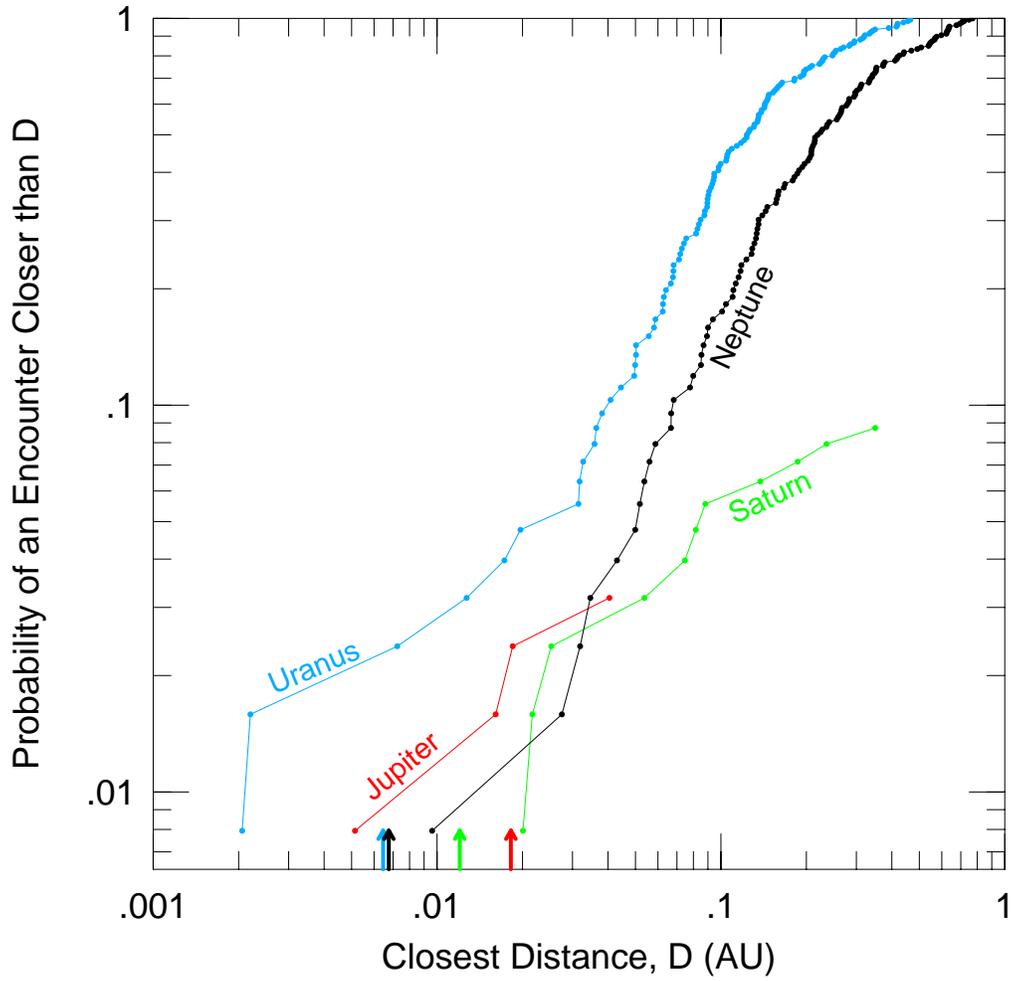}
\caption{The cumulative probability of encounters at a given distance with each of the giant planets.  The arrows along the x-axis show the approach distance, D, for which binary disruption becomes probable in a single encounter.  As shown by the curves, Jupiter and Uranus are the most likely to disrupt a \cR-like object.  }
\label{fig_close} 
\end{figure}

\newpage\begin{table*}
\label{orbit}
\begin{tabular}{lcccccccr}
\hline
				& q 		&  a 		& e		& i 		& H$_V$		& limit			& camera 		\\ 			
Object			& (AU)	&  (AU)	& 		& (deg)	& (mag)$^1$	& ($\Delta$ mag)$^2$& 	 		\\ 
\noalign { \vskip 12pt \hrule height 1pt \vskip 1pt \hrule height 1pt \vskip 8pt } 
{\bf Binary}		&		&		&		&		&			&				&			\\
\CR				& 17.5	& 38.1	& 0.54	& 2.4 	& 7.65(1)$^3$	& 7.4			& HRC 		\\   	
{\bf Single}		&		&		&		&		&			&				&			\\
(49036) Pelion 		& 17.2	& 19.9	& 0.14	& 9.4	& 10.54(2)$^4$	& 5.3			& HRC		\\ 	
(54598) Bienor 		& 13.2	& 16.5	& 0.20	& 20.8	& 7.69(1)$^3$	& 8.1			& HRC		\\ 	
\TD				& 12.3	& 95.1	& 0.87	& 6.0	& 9.06(5)$^3$	& 7.6			& STIS		\\ 
\BU 				& 20.6	& 33.4	& 0.38	& 14.2	& 7.2(1) $^5$	& 5.9			& STIS		\\
\PJ				& 28.6	& 124	& 0.77	& 5.7	& 8.0(1)$^5$	& 1.3			& NICMOS	\\
\EE				& 22.6	& 49.8	& 0.55	& 5.9	& 8.49(1)$^3$	& 4.3			& WFPC2		\\ 	
\OO				& 20.8	& 524	& 0.96	& 20.1	& 9.8(1)$^3$	& 3.0			& WFPC2		\\ 
\noalign { \vskip 12pt \hrule height 1pt \vskip 8pt } 
\end{tabular} 
\caption[]{{\bf Centaurs Observed with HST} \\
 $^1$ Uncertainty in final digit shown in parentheses. $^2$ Detection limit for faint companions $= 2.5$ log $(p/3\sigma$) where $p$= peak pixel in the source and $\sigma$ is the rms background variation in the combined (multidrizzled) image.  For \CR\ the peak pixel refers to the primary.  $^3$ Tegler et al.~2003; $^4$ Tegler and Romanishin 2000;  $^5$ Minor Planet Center
 }
\end{table*}

\end{document}